\documentclass[a4, journal]{IEEEtran}
\IEEEoverridecommandlockouts
\usepackage{cite}
\usepackage[hyphens]{url}

\usepackage{amsmath,amssymb,amsfonts}

\usepackage{graphicx}
\usepackage{textcomp}
\usepackage{xcolor}
\usepackage{subfigure}
\usepackage{multirow}
\usepackage{tabularx}
\usepackage{gensymb}
\usepackage{float}
\usepackage{svg}
\usepackage{optidef}
\usepackage[ruled,vlined]{algorithm2e}
\usepackage{hyperref}
\newcommand{\linebreakand}{%
  \end{@IEEEauthorhalign}
  \hfill\mbox{}\par
  \mbox{}\hfill\begin{@IEEEauthorhalign}
}
\usepackage{nomencl}
\usepackage{etoolbox}
\def\BibTeX{{\rm B\kern-.05em{\sc i\kern-.025em b}\kern-.08em
    T\kern-.1667em\lower.7ex\hbox{E}\kern-.125emX}}
    
\renewcommand\nomgroup[1]{%
  \item[\bfseries
  \ifstrequal{#1}{A}{Sets}{%
  \ifstrequal{#1}{B}{Loads and Generation}{%
  \ifstrequal{#1}{C}{Storage}{%
  \ifstrequal{#1}{D}{Investment}{}}}}%
]}

\makenomenclature
\begin{document}

\title{Optimizing Deep Decarbonization Pathways in California with Power System Planning Using Surrogate Level-based Lagrangian Relaxation}

\author{Osten~Anderson,~\IEEEmembership{Student Member,~IEEE,}
        Mikhail~A.~Bragin,~\IEEEmembership{Senior~Member,~IEEE,}
        and Nanpeng~Yu,~\IEEEmembership{Senior~Member,~IEEE}
}
\maketitle
\IEEEpubidadjcol

\begin{abstract}
With California's ambitious goals to decarbonize the electrical grid by 2045, significant challenges arise in power system investment planning. Existing modeling methods and software focus on computational efficiency, which is currently achieved by simplifying the associated unit commitment formulation. This may lead to unjustifiable inaccuracies in the cost and constraints of gas-fired generation operations and may affect both the timing and the extent of investment in new resources, such as renewable energy and energy storage. 
To address this issue, this paper develops a more detailed and rigorous mixed-integer programming model, and more importantly, a solution methodology utilizing surrogate level-based Lagrangian relaxation to address the combinatorial complexity that results from the enhanced level of model detail. This allows us to optimize a model with approximately 12 million binary and 100 million total variables in under 48 hours. 
The investment plan is compared with those produced by E3's RESOLVE software, which is currently employed by the California Energy Commission and California Public Utilities Commission. Our model produces an investment plan that differs substantially from that of the existing method and saves California over 4 billion dollars over the investment horizon. 
\end{abstract}


\printnomenclature

\section{Introduction}
California's Senate Bill 100 (SB100) mandates that all retail electrical sales come from non-carbon sources by 2045. Consequently, a comprehensive investment plan is essential to identify the most cost-effective approach to achieve this goal. Over the approximately 20-year planning horizon, operating costs are expected to reach hundreds of billions of dollars. As a result, even a small percentage improvement in the total investment and operation cost for the decarbonization plan could yield savings of hundreds of millions of dollars in investment and operational expenses. To this end, appropriate modeling and solution methodologies capable of handling such massive decision-making problems are crucial.

From a modeling perspective, the planning problem is frequently modeled as a mixed-integer linear programming (MILP) problem to exploit the capability of existing MILP solvers such as Gurobi \cite{gurobi}, Xpress \cite{xpress}, and CPLEX \cite{cplex}. Over the 20-year decarbonization planning horizon, decisions regarding resource dispatch are co-optimized with decisions associated with the construction and the retirement of energy resources. While many renewable and storage resources can be modeled using continuous variables, thermal unit behavior (commitment, decommission, and investment) can only be accurately captured with binary variables. This requirement, combined with the extensive time horizon, leads to an issue known as combinatorial complexity - as the planning horizon increases linearly, the associated complexity increases superlinearly (e.g., exponentially). Consequently, when using off-the-shelf commercial software, large MILP planning problems can quickly become increasingly challenging to solve, with no guarantee that even a feasible solution can be found within a reasonable CPU time.  

While LP-relaxed problems may still theoretically be NP-hard, the practical performance of such methods as simplex or barrier methods within commercial software leads to much-reduced CPU times. However, the LP-relaxed version of unit commitment cannot accurately capture the behavior of thermal units and tends to overestimate their operational flexibility. Consequently, an investment plan based on these simplifications may lead to higher costs or even reliability issues when subjected to the constraints of real-world operations. 

In this paper, we address the California decarbonization planning issue by formulating it as a MILP problem. This approach provides a more accurate model of thermal plants' operations as compared to previous simplifications. Instead of oversimplifying the model, we tailor a surrogate Lagrangian relaxation technique to decompose the problem into manageable subproblems. This method significantly reduces the combinatorial complexity and uses Lagrangian multipliers for iterative coordination of the subproblems. By using the proposed method, investment plans are more consistent with real-world power system operations.
The results are compared with RESOLVE, a model used by California state agencies for decarbonization studies. We find that RESOLVE routinely underestimates the investment required to meet intermediate emissions targets. Further, our model results in lower overall costs on the order of billions through the investment horizon. 

The remainder of the paper is structured as follows. Section \ref{sec2} discusses related works in optimization and power system planning. Section \ref{sec3} formulates the optimization model and solution methodology. Section \ref{sec4} presents numerical testing results as well as comparisons to the existing linearized model results. 
Section \ref{sec5} concludes the paper. 

\section{Related Work} \label{sec2}
While the decarbonization planning problem is relatively new, it is closely connected to the generation expansion problem, which has been the subject of study for decades. Further, as the focus shifts to renewable generation resources, the lines between these problems have blurred in the literature. This section will review related literature in both decarbonization and generation expansion planning (GEP), as well as works related to the proposed surrogate Lagrangian relaxation solution methodology. 

\subsection{Generation Expansion Planning}
The decarbonization planning problem is essentially a modification of the generation expansion problem, with more emphasis on the construction of green technology and subject to constraints on emissions. Thus, these problems will be considered together. GEP approaches generally fall into two groups: reduction in model complexity and alternative optimization methods. Reduction in model complexity refers to relaxations made to the full MILP unit commitment formulation used within GEP. Within this group, there are a few common streams. The first is the relaxation of integrality requirements within unit commitment \cite{HULL1, HULL2, GEP1}. The general drawback to methods of this class is an overestimation of the flexibility of thermal units and pumped storage. The second is the omission of detailed technical constraints within the unit commitment formulation. 
For example, these methods may model clusters rather than individual thermal units \cite{GEP6} or may not model thermal unit operational constraints like ramping \cite{GEP4}. 
Simpler models may even neglect temporal dependencies and instead model using a load duration curve or similar metrics \cite{GEP7, GEP5, GEP3}.
In general, even if the fleet may be able to satisfy a forecasted peak load, it may not satisfy such load subject to more granular operational constraints on thermal units, and provide all necessary reliability products. On the other hand, not modeling the full detail of unit commitment may lead to a solution that is feasible, but much more expensive, during realistic operation, due to the need to satisfy more granular operational constraints.

Heuristic optimization methods are those that do not rely purely on traditional optimization software, like CPLEX or Gurobi. Notable examples of these methods include genetic algorithm \cite{GA1} and particle swarm optimization \cite{PSO1, PSO2}.
One of the shortcomings of these methods is the lack of a lower bound to provide a measure of the solution quality - how close a solution is to the global optimal. Moreover, heuristic and metaheuristics methods such as particle swarm optimization may generally suffer from getting stuck in local optima.

Another approach to improving computational tractability is the use of Benders' decomposition. This method has been deployed for transmission expansion planning in many studies \cite{bender_tep2, bender_tep3, bender_tep4}. In \cite{bender_gep}, Benders' decomposition is applied to generation expansion planning, however, it still relies on a heuristic, genetic algorithm, and thus inherits those intrinsic issues. In general, Benders' decomposition has been applied to generation expansion much less than transmission expansion. 

Several open-source generation expansion packages have been developed, allowing users to solve generation or transmission expansion planning without performing modeling themselves. These include Gridpath \cite{gridpath}, GenX \cite{genx}, and ReEDS \cite{reeds}.
These packages vary in terms of solving MILP vs LP, clustered vs non-clustered, and level of unit commitment detail. Some even allow users to choose which of these detail levels is used in unit commitment. 
However, even where these packages permit generator-level integer unit commitment modeling, no steps are taken to improve computational tractability therein, and thus are severely compromised in their ability to be applied to any significant number of time periods. 

This work closely follows Energy + Environmental Economics' (E3) RESOLVE model's data and general composition \cite{RESOLVE}. The RESOLVE model is used by California Public Utilities Commission to perform integrated resource planning in order to meet California's long-term energy policy goals.
However, the RESOLVE model presents a linearized, clustered version of unit commitment, in which units are clustered by similar technology and scheduled linearly within these clusters. This simplification massively increases computational efficiency, but has the potential to overestimate the flexibility of thermal units. Clustered unit commitment cannot distinguish between units for tracking constraints like minimum uptime. Neglecting linear variables leads to physical impossibilities, like turning on half of a unit. This overestimated operational flexibility will likely lead to an investment solution that is sub-optimal when the full set of constraints are applied in real-world operations. In this work, we overcome these issues by presenting a more detailed operational model.

\subsection{Optimization Methods: Lagrangian Relaxation} 
Lagrangian relaxation (LR) has been a powerful technique to solve MIP problems by exploiting the drastic reduction of complexity and has been especially efficient for solving decomposable problems in power systems like unit commitment \cite{sun2018novel, wu2021novel} and beyond \cite{bragin2015convergence, bragin2018scalable, bragin2022surrogate}. LR can frequently achieve performance enhancements over commercially available software by a factor of 2-3 times, and in some cases, improve the CPU time by multiple orders of magnitude. Lagrangian relaxation is therefore expected to be useful for power system investment planning as well. Key features of LR include the decomposition of the problem into subproblems and the exponential reduction of complexity, enabling efficient coordination through iterative updates of multipliers. However, traditional LR faced difficulties in updating multipliers because of the high effort required to obtain subgradient direction; even if obtained, subgradient directions tend to change drastically and lead to zigzagging of multipliers and slow convergence. Additionally, standard LR requires the optimal dual value knowledge for convergence as in Polyak stepsizing \cite{polyak1969minimization}.

Recently, surrogate Lagrangian relaxation (SLR) \cite{bragin2015convergence} has addressed most of these issues, enabling multiplier updates with only one ``good-enough'' subproblem solution at a time that satisfies the ``surrogate optimality condition''. This procedure essentially improves the incumbent solution of a relaxed problem (rather than finding the exact optimal solution) in a computationally efficient way due to the drastic reduction of complexity, while still guaranteeing convergence, and reducing the zigzagging of multipliers. From the subproblem-coordination standpoint, the method eliminates the need for optimal dual value knowledge.  Moreover, several advancements have been made to the SLR framework, including surrogate absolute-value Lagrangian relaxation (SAVLR) \cite{bragin2018scalable} which accelerates convergence through piece-wise penalties.
These methods have demonstrated success in solving various complex problems.

Furthermore, the surrogate ``level-based" Lagrangian relaxation (SLBLR) technique has been developed \cite{bragin2022surrogate}, which uses the Polyak stepsize formula with efficient level-value determination without requiring estimation or heuristic adjustments. This user-friendly approach is robust and reduces the need for domain knowledge. SLBLR has successfully addressed major issues of previous methods, specifically obviating the need to know the optional function value while still exploiting the geometric (fastest possible) convergence inherent in the Polyak formula. This makes it suitable for coordinating multiple subsystems and supporting complex problem decision-making, such as in large-scale power system investment planning for California's decarbonization goals.

\section{Technical Method} \label{sec3}
\nomenclature[A, 01]{$t, T$}{Index, set of hour}
\nomenclature[A, 01]{$w, W$}{Index, set of week}
\nomenclature[A, 01]{$y, Y$}{Index, set of year}
\nomenclature[A, 02]{$u, U$}{Index, set of thermal unit}
\nomenclature[A, 03]{$s, S$}{Index, set of storage resource}
\nomenclature[A, 04]{$r, R$}{Index, set of renewable resource}
\nomenclature[A, 05]{$h, H$}{Index, set of large hydro resource}
\nomenclature[A, 06]{$z, Z$}{Index, set of balancing authority zone}
\nomenclature[A, 07]{$l, L$}{Index, set of line}

\nomenclature[A, 15]{$U_z$}{Subset of thermal resources in zone $z$}
\nomenclature[A, 16]{$S_z$}{Subset of storage resources in zone $z$}
\nomenclature[A, 17]{$R_z$}{Subset of renewable resources in zone $z$}
\nomenclature[A, 18]{$H_z$}{Subset of large hydro resources in zone $z$}

\nomenclature[B, 01]{$\mathcal{L}_z(t)$}{Load in zone $z$ at time $t$ (MW)}
\nomenclature[B, 02]{$v_{u}(t)$}{On/off status of unit $u$ at time $t$ (1, 0) }
\nomenclature[B, 03]{$p_{u}(t)$}{Power output of unit $u$ at time $t$ (MW)}
\nomenclature[B, 04]{$p_r(t)$}{Power output of renewable resource $r$ at time $t$ (MW)}
\nomenclature[B, 05]{$p_h(t)$}{Power output of large hydro resource $h$ at time $t$ (MW)}
\nomenclature[B, 06]{$UT_u$}{Minimum uptime of unit $u$ (hours)}
\nomenclature[B, 07]{$DT_u$}{Minimum downtime of unit $u$ (hours)}
\nomenclature[B, 08]{$RU_u$}{Ramp up rate of unit $u$ (MW/hour)}
\nomenclature[B, 09]{$RD_u$}{Ramp down rate of unit $u$ (MW/hour)}
\nomenclature[B, 10]{$SU_u$}{Startup power limit of unit $u$ (MW)}
\nomenclature[B, 11]{$SD_u$}{Shutdown power limit of unit $u$ (MW)}
\nomenclature[B, 12]{$\underline{P}_u$}{Minimum output of unit $u$ (MW)}
\nomenclature[B, 13]{$\overline{P}_u$}{Maximum output of unit $u$ (MW)}
\nomenclature[B, 14]{$\underline{P}_h$}{Minimum output of hydro resource $h$ (MW)}
\nomenclature[B, 15]{$\overline{P}_h$}{Maximum output of hydro resource $h$ (MW)}
\nomenclature[B, 16]{$RL_h$}{Ramping limit of hydro resource $h$ (MW/hour)}
\nomenclature[B, 17]{$B_h$}{Weekly energy budget of hydro resource $h$ (MWh)}
\nomenclature[B, 18]{$f_l(t)$}{Flow on line $l$ at time $t$ (MW)}
\nomenclature[B, 18.5]{$\lambda_{l, z}$}{Incidency of line $l$ on zone $z$}
\nomenclature[B, 19]{$\underline{F}_l$}{Minimum (negative) flow on line $l$ (MW)}
\nomenclature[B, 19.5]{$\overline{F}_l$}{Maximum flow on line $l$ (MW)}
\nomenclature[B, 20]{$IC_r$}{Installed capacity of renewable resource $r$ (MW)}
\nomenclature[B, 21]{$PF_r(t)$}{Production factor of renewable resource $r$ at time $t$}
\nomenclature[B, 22]{$p_r^{curt}(t)$}{Curtailment of renewable resource $r$ at time $t$ (MW)}
\nomenclature[B, 23]{${c^{curt}_r}$}{Cost of curtailment of resource $r$ (\$/MWh)}

\nomenclature[B, 24]{$SUC_u(t)$}{Startup cost of unit $u$ at time $t$ (\$)}
\nomenclature[B, 25]{$SDC_u(t)$}{Shutdown cost of unit $u$ at time $t$ (\$)}
\nomenclature[B, 26]{$GCS_u$}{Generation cost slope of unit $u$ (\$/MWh)}
\nomenclature[B, 27]{$GCI_u$}{Generation cost intercept of unit $u$ (\$/hour)}

\nomenclature[C, 01]{$v_s(t)$}{Storage charge (0)/discharge (1) status at time $t$}
\nomenclature[C, 02]{$p_{s}^c(t)$}{Storage rate of charge at time $t$ (MW)}
\nomenclature[C, 03]{$p_s^d(t)$}{Storage rate of discharge at time $t$ (MW)}
\nomenclature[C, 04]{$\overline{p}_s^{c}$}{Storage max rate of charge (MW)}
\nomenclature[C, 05]{$\overline{p}_s^{d}$}{Storage max rate of discharge (MW)}
\nomenclature[C, 06]{$\overline{C}_s$}{Storage max state of charge (MWh)}
\nomenclature[C, 07]{$\underline{C}_s$}{Storage min state of charge (MWh)}
\nomenclature[C, 08]{$C_s(t)$}{Storage state of charge at time $t$ (MWh)}
\nomenclature[C, 09]{$\eta_s^c$}{Storage charge efficiency}
\nomenclature[C, 10]{$\eta_s^d$}{Storage discharge efficiency}
\nomenclature[C, 11]{$\delta_s$}{Storage self discharge}

\nomenclature[D, 01]{$IU_u(y)$}{Install status of unit $u$ in year $y$}
\nomenclature[D, 02]{$IU^p_u(y)$}{Planned install status of unit $u$ in year $y$}
\nomenclature[D, 03]{$IU^b_u(y)$}{Build flag for unit $u$ in year $y$}
\nomenclature[D, 04]{$IU^r_u(y)$}{Retirement flag for unit $u$ in year $y$}
\nomenclature[D, 05]{$IC_s(y)$}{Installed capacity of storage resource $s$ in year $y$}
\nomenclature[D, 06]{$IC^p_s(y)$}{Planned capacity of storage resource $s$ in year $y$}
\nomenclature[D, 07]{$IC^b_s(y)$}{Built capacity of storage resource $s$ in year $y$}
\nomenclature[D, 08]{$ICE_s(y)$}{Installed energy capacity of storage resource $s$ in year $y$}
\nomenclature[D, 09]{$ICE^p_s(y)$}{Planned energy capacity of storage resource $s$ in year $y$}
\nomenclature[D, 10]{$ICE^b_s(y)$}{Built energy capacity of storage resource $s$ in year $y$}

\nomenclature[D, 11]{$IC_r(y)$}{Installed capacity of renewable resource $r$ in year $y$}
\nomenclature[D, 12]{$IC^p_r(y)$}{Planned capacity of renewable resource $r$ in year $y$}
\nomenclature[D, 13]{$IC^b_r(y)$}{Built capacity of renewable resource $r$ in year $y$}

\nomenclature[D, 14]{$C_y^{gen}$}{Generation costs in year  $y$}
\nomenclature[D, 15]{$C_y^m$}{Maintenance costs in year  $y$}
\nomenclature[D, 16]{$C_y^{inv}$}{Investment costs in year  $y$}

In this section, we present the planning problem as a two-timescale optimization formulation. At the hourly level, we formulate a unit commitment problem, responsible for dispatching resources to meet load and ancillary service requirements. Meanwhile, the yearly level modeling focuses on investment decisions, governing the construction or retirement of resources and their corresponding timelines. Subsection \ref{sec3A} presents the formulation for single-week unit commitment. Subsection \ref{sec3B} then integrates the single-week unit commitment formulation into the multi-week, multi-year planning model.
    
\subsection{Single-Week Unit Commitment} \label{sec3A}

Unit commitment (UC) will be considered over a time period $T$ with 1-hour resolution. 
Every UC variable is indexed temporally by $(y, w, t)$, a tuple of year, week, and hour. However, for this subsection, we can formulate all constraints for an arbitrary year and week. Thus, we will hide the $y, w$ indices and let $(y, w, t) \rightarrow (t)$. In the planning model formulation, these constraints will be enforced over all $y \in Y, w \in W$.
To circumvent the need to define the initial status of units in UC, time periods will be considered to be circular in a fashion echoing that of the RESOLVE package \cite{RESOLVEcode}. That is, all constraints that link hours are enforced between the end of the period back to the beginning. 
\begin{figure}
\centering
\includegraphics[width=0.75\linewidth]{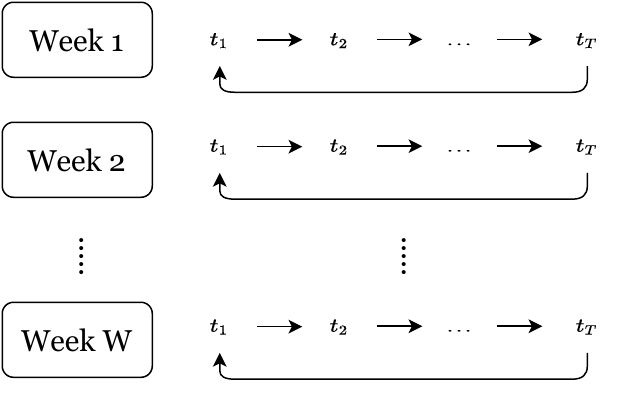}
\caption{Visualization of the circular representation of time.}
\label{fig:circulartime}
\end{figure}

The circular representation of time can be expressed via the modulo operator. A negative argument $t$ in $\tau(t)$ maps backward from the end of the period. For constraints looking past the hourly optimization window, time maps back to the beginning of the window,  as shown in Fig. \ref{fig:circulartime}. Only $t$ is considered circular, not $y$ or $w$.
For the regular period, $t \in [1, T]$, $t=\tau(t)$ and $\tau$ can be omitted for notational brevity. 
\begin{equation}
    \tau(t) = \mod(t-1+T, T)
\end{equation}
\subsubsection{Generation Resources}
The generation fleet consists of five basic types of generation resources: thermal units, renewable resources, firm resources, storage resources, and large hydro resources. These formulations are based on the formulations presented in \cite{RESOLVEcode}, \cite{batteryUC}, and \cite{UC}. These types of resources and their constraints will now be discussed. 

\noindent \textbf{Thermal Units.} Thermal units comprise various types of gas-fired power plants, including combined-cycle gas turbines, peakers, steam turbines, aeroderivative combustion turbines, and coal-fired plants. Within each zone, each technology type is represented by one or two ``typical'' units characterized by their physical parameters, with every unit of that technology sharing the same physical parameters, like capacity and ramp rates. The output of thermal units are subject to the following constraints: minimum and maximum output \eqref{gen1}, minimum uptime and downtime \eqref{uptime}, \eqref{downtime}, and ramping limits \eqref{rampup}, \eqref{SDramp}, \eqref{rampdown}.

\begin{equation} \label{gen1}
\begin{split}
    \underline{P}_uv_u(t) \leq p_u(t) \leq \overline{P}_uv_u(t), \; \forall t \in T, u \in U
\end{split}
\end{equation}
\begin{equation} \label{uptime}
\begin{split}
    \sum^{\tau(t+UT_u)}_{n=t+1} v_u(n) \geq UT_u [ v_u(t)-v_u(t-1)],\\
    \forall t \in T, u \in U
\end{split}
\end{equation}
\begin{equation} \label{downtime}
\begin{split}
    \sum^{\tau(t+DT_u)}_{n=t+1} [1-v_u(n)] \geq DT_u [ v_u(t-1)-v_u(t)],\\
    \forall t \in T, u \in U
\end{split}
\end{equation}
\begin{equation} \label{rampup}
\begin{split}
    p_u(t) \leq  p_u(\tau(t-1)) + RU_uv_u(\tau(t-1))  \\
    + SU_u[v_u(t)-v_u(\tau(t-1))]+\overline{P}_u(1-v_u(t)),\\
    \forall t \in T, u \in U
\end{split}
\end{equation}
\begin{equation}
\begin{split} \label{SDramp}
p_u(t) \leq \overline{P}_uv_u(\tau(t+1)) +  SD_u [v_u(t)-v_u(\tau(t+1))], \\
\forall t \in T, u \in U
\end{split}
\end{equation}
\begin{equation} \label{rampdown}
\begin{split}
    p_u(t) \geq p_u(\tau(t-1)) - RD_uv_u(t)  \\  - SD_u[v_u(\tau(t-1))-v_u(t)]-\overline{P}_u[1-v_u(\tau(t-1))], \\
    \forall t \in T, u \in U
\end{split}    
\end{equation} 

Thermal units are associated with fuel costs, which are modeled as linear with $p_u(t)$ and $v_u(t)$, and startup/shutdown costs, which are assessed a fixed cost $suc_u$/$sdc_u$ whenever the unit turns on/off.
\begin{equation}
\begin{split}
        SUC_u(t) = max(0, v_u(t)-v_u(\tau(t-1))) \cdot suc_u, \\ \forall t \in T, u \in U
\end{split}
\end{equation}
\begin{equation}
\begin{split}
    SDC_u(t) = max(0, v_u(\tau(t-1))-v_u(t)) \cdot sdc_u, \\ \forall t \in T, u \in U
\end{split}
\end{equation}

\noindent \textbf{Renewable Resources.} Renewable resources are utility-scale solar and wind farms, and an agglomeration of behind-the-meter customer solar. Renewable resources can be described by a generation shape. Each resource generates a certain percentage of its total rated capacity $PF_r(t)$  in a given hour, depending on the solar irradiance or wind speed at its location. The power output of a renewable resource \eqref{renpower} is equal to this amount minus any curtailment. Curtailment is associated with a cost related to the loss of production tax credits, ${c}^{curt}_r$.
\begin{equation} \label{renpower}
    p_r(t) = IC_r \cdot PF_r(t) - p_r^{curt}(t), \; \forall t \in T, r \in R
\end{equation}

\noindent \textbf{Firm resources}. Firm resources, which include nuclear, small hydro, biofuel, geothermal, and combined heat and power, will be lumped with renewable resources. Firm resources produce a fixed output every hour and are not schedulable or curtailable, so for firm resources within the set of renewable resources $R$, $p_r^{curt}(t) = 0, t \in T$. Firm generation can, however, vary by season, such as due to maintenance for nuclear or for flow rate changes in small hydro.

\noindent \textbf{Large Hydro Units.} Large hydro units are dispatchable hydro resources, which are subject to weekly energy budget constraints \eqref{Hydro energy budget}, ramp limits \eqref{Hydro ramp limits}, and generation capacity constraints \eqref{Hydro gen limits}.
\begin{equation} \label{Hydro energy budget}
    \sum_{t \in T} p_h(t) \times 1 \:  hour \leq B_h, \; \forall h \in H
\end{equation}
\begin{equation} \label{Hydro ramp limits}
\begin{split}
    p_h(t)-RL_h \leq p_h(\tau(t+1)) \leq p_h(t)+RL_h, \\ \forall t \in T, h \in H
\end{split}
\end{equation}
\begin{equation} \label{Hydro gen limits}
\begin{split}
    \underline{P}_h \leq p_h(t) \leq \overline{P}_h,\; \forall t \in T, h \in H
\end{split}
\end{equation}

\noindent \textbf{Storage Resources.} Storage resources include pumped and battery storage. Storage resources are defined by their power rating (MW) and energy rating (MWh). Storage resources can charge using overgeneration and discharge to serve undergeneration. Storage resources are limited by their rated power and energy capacity. To enforce minimum duration, particularly for pumped storage resources, storage resources have a binary discharge (1) or charge (0) status. Charge and discharge rates are modeled separately to account for efficiency losses and are subject to minimum and maximum rate constraints (\ref{charge}, \ref{discharge}). 
\begin{equation} \label{charge}
\begin{split}
    0 \leq p_{s}^c(t) \leq (1-v_{s}) \overline{p}_{s}^{c}, \; \forall t \in T, s \in S
\end{split}
\end{equation}
\begin{equation} \label{discharge}
\begin{split}
    0 \leq p_{s}^d(t) \leq v_{s} \overline{p}_{s}^{d}, \; \forall  t \in T, s \in S
\end{split}
\end{equation}

Storage resources are subject to battery capacity limits - a minimum and maximum state of charge constraint \eqref{capacity}:
\begin{equation} \label{capacity}
\begin{split}
    \underline{C}_{s} \leq C_{s}(t) \leq \overline{C}_{s}, \; \forall t \in T, s \in S.
\end{split}
\end{equation}
Storage resource state of charge balance is governed by \eqref{SoC}. 
\begin{multline} \label{SoC}
C_{s}(t) = (1-\delta_{s}) C_{s}(\tau(t-1)) \\
 + \left[(1-v_{s}) p_{s}^c(t) \eta_{s}^c - v_{s} p_{s}^d(t) \frac{1}{\eta_{s}^d}\right] \times 1 \: hour,
\\  \forall t \in T, s \in S
\end{multline}

\subsubsection{Zones and Lines}
A zonal unit commitment model is employed to represent the California Independent System Operator (CAISO) and the Western Interconnection, encompassing distinct zones: CAISO, three balancing authorities within California (LADWP, IID, BANC), and two out-of-state aggregations (NW, SW). Each zone is interconnected with at least one other zone through transmission lines, enabling power transfer between zones as a decision variable. This approach omits the need for detailed power flow analyses, reducing the computational complexity of the problem while facilitating a comprehensive representation of the interconnected system. The incidency of line $l$ on zone $z$ is captured by $\lambda_{l,z}$, where a value of 0 denotes non-incidence and a value of 1 and -1 denote reference directions of line $l$ into and out of zone $z$, respectively. Transmission is associated with a transmission cost $c_l^{tx}$ which captures wheeling costs, and can be derived from Open
Access Transmission Tariffs \cite{OATT}. The power flows are subject to line capacity constraints: 
\begin{equation} \label{Linelimits}
\begin{split}
    \underline{F}_l \leq f_l(t) \leq \overline{F}_l, \; \forall t \in T, l \in L.
\end{split}
\end{equation}
\subsubsection{Load and Reserve Requirements} 
The ancillary service requirements must be satisfied with resources in CAISO.
Each reserve product is modeled individually. Superscripts $fr$, $sr$, $lf\uparrow$, $reg\uparrow$ denote frequency response, spinning reserve, load following up, and regulation up products, while $lf\downarrow$, $reg\downarrow$ denote the load following down and regulation down products. Products can only be supplied up to the headroom and footroom available for thermal units \eqref{thermalhead} \eqref{thermalfoot} and hydro units \eqref{hydrohead} \eqref{hydrofoot}. 
Thermal provision of frequency response is limited to 8\% of the current output \eqref{frdispatch}. 
For products other than frequency response, ramping limits must also be obeyed for thermal units (\ref{thermalrampup}) (\ref{thermalrampdown}) and hydro units (\ref{hydrorampup}) (\ref{hydrorampdown}). Storage can provide each product up to the headroom and footroom of both power capacity \eqref{storagepowerhead} \eqref{storagepowerfoot} and energy capacity \eqref{storageenergyhead} \eqref{storageenergyfoot}. Up to half of the load following down can be provided via curtailable renewable resources \eqref{rncurtlf} up to the available footroom \eqref{rncurtlffoot}. 

\begin{equation} \label{thermalhead}
\begin{split}
    p_u^{fr}(t) + p_u^{sr}(t) + p_u^{lf\uparrow}(t) + p_u^{reg\uparrow}(t) \\ <= \overline{P}_uv_u(t) - p_u(t), \; \forall t \in T, u \in U
\end{split}
\end{equation}
\begin{equation} \label{thermalfoot}
    p_u^{lf\downarrow}(t) + p_u^{reg\downarrow}(t) <= p_u(t) - \underline{P}_uv_u(t), \; \forall t \in T, u \in U 
\end{equation}
\begin{equation} \label{frdispatch}
    p_u^{fr}(t) <= 0.08 p_u(t), \; \forall t \in T, u \in U
\end{equation}
\begin{equation} \label{thermalrampup}
    p_u^{sr}(t) + p_u^{lf\uparrow}(t) + p_u^{reg\uparrow}(t) <= RU_u/6, \; \forall t \in T, u \in U 
\end{equation}
\begin{equation} \label{thermalrampdown}
    p_u^{lf\downarrow}(t) + p_u^{reg\downarrow}(t) <= RD_u/6, \; \forall t \in T, u \in U
\end{equation}
\begin{equation} \label{hydrohead}
\begin{split}
    p_h^{fr}(t) + p_h^{sr}(t) + p_h^{lf\uparrow}(t) + p_h^{reg\uparrow}(t) \\  <= \overline{P}_h - p_h(t), \; \forall t \in T, h \in H
\end{split}
\end{equation}
\begin{equation} \label{hydrofoot}
    p_h^{lf\downarrow}(t) + p_h^{reg\downarrow}(t) <= p_h(t) - \underline{P}_h, \; \forall t \in T, h \in H 
\end{equation}
\begin{equation}  \label{hydrorampup}
    p_h^{sr}(t) + p_h^{lf\uparrow}(t) + p_h^{reg\uparrow}(t) <= RL_h/6, \; \forall t \in T, h \in H 
\end{equation}
\begin{equation} \label{hydrorampdown}
    p_h^{lf\downarrow}(t) + p_h^{reg\downarrow}(t) <= RL_h/6, \; \forall t \in T, h \in H 
\end{equation}
\begin{equation}
\begin{split} \label{storagepowerhead}
    p_s^{fr}(t) + p_s^{sr}(t) + p_s^{lf\uparrow}(t) + p_s^{reg\uparrow}(t) \\ <= \overline{p}^d_s(t) - p^d_s(t) + p^c_s(t), \; \forall t \in T, s \in S
\end{split}
\end{equation}
\begin{equation} \label{storagepowerfoot}
\begin{split}
    p_s^{lf\downarrow}(t) + p_s^{reg\downarrow}(t) <= \overline{p}^c_s(t) - p^c_s(t) + p^d_s(t), \\ \forall t \in T, s \in S
\end{split}
\end{equation}
\begin{equation} \label{storageenergyhead}
\begin{split}
    p_s^{fr}(t) + p_s^{sr}(t) + p_s^{lf\uparrow}(t) + p_s^{reg\uparrow}(t) <= C_s(t) - \underline{C}_s, \\  \forall t \in T, s \in S
\end{split}
\end{equation}
\begin{equation} \label{storageenergyfoot}
    p_s^{lf\downarrow}(t) + p_s^{reg\downarrow}(t) <=\overline{C}_s - C_s(t), \; \forall t \in T, s \in S 
\end{equation}
\begin{equation} \label{rncurtlf}
    p_r^{lf\downarrow}(t) <= 0.5 LF\downarrow(t), \; \forall t \in T, r \in R 
\end{equation}
\begin{equation} \label{rncurtlffoot}
\begin{split}
    p_r^{lf\downarrow}(t) <= IC_r \cdot PF_r(t) - p_r^{curt}(t) - p_r(t), \\  \forall t \in T, r \in R
\end{split}
\end{equation}

770MW must be held at all times for frequency regulation. Regulation up, regulation down, and spinning reserve each require 1\% of the CAISO load. Load following up and down requirements are based upon renewable penetration scenario analysis carried out by E3. 
\begin{equation}
\begin{split}
     \sum_{u \in U_z} p_u^{fr}(t) + \sum_{h \in H_z} p_h^{fr}(t) + \sum_{s \in S_z} p_s^{fr}(t) \\  >= 770MW, \; t \in T, z=0
\end{split}
\end{equation}
\begin{equation}
\begin{split}
    \sum_{u \in U_z} p_u^{sr}(t) + \sum_{h \in H_z} p_h^{sr}(t) + \sum_{s \in S_z} p_s^{sr}(t) \\  >= 0.01 \mathcal{L}_z(t), \; t \in T, z=0
\end{split}
\end{equation}
\begin{equation}
\begin{split}
     \sum_{u \in U_z} p_u^{reg\uparrow}(t) + \sum_{h \in H_z} p_h^{reg\uparrow}(t) + \sum_{s \in S_z} p_s^{reg\uparrow}(t) \\  >= 0.01 \mathcal{L}_z(t), \; t \in T, z=0
\end{split}
\end{equation}
\begin{equation}
\begin{split}
    \sum_{u \in U_z} p_u^{reg\downarrow}(t) + \sum_{h \in H_z} p_h^{reg\downarrow}(t) + \sum_{s \in S_z} p_s^{reg\downarrow}(t) \\  >= 0.01 \mathcal{L}_z(t), \; t \in T, z=0
\end{split}
\end{equation}
\begin{equation}
\begin{split}
    \sum_{u \in U_z} p_u^{lf\uparrow}(t) + \sum_{h \in H_z} p_h^{lf\uparrow}(t) + \sum_{s \in S_z} p_s^{lf\uparrow}(t) \\  >= LF\uparrow(t), \; t \in T, z=0
\end{split}
\end{equation}
\begin{equation} \label{lastreservereq}
\begin{split}
    \sum_{u \in U_z} p_u^{lf\downarrow}(t) + \sum_{h \in H_z} p_h^{lf\downarrow}(t) + \sum_{s \in S_z} p_s^{lf\downarrow}(t) \\  + \sum_{r \in R_z} p_r^{lf\downarrow}(t)  >= LF\downarrow(t), \; t \in T, z=0
\end{split}
\end{equation}

Each zone within the ISO area must satisfy the zonal power balance constraints \eqref{Nodalpowerbalance} as:
\begin{multline} \label{Nodalpowerbalance}
\sum_{u \in U_z} p_u(t) + \sum_{s \in S_z}[p_{s}^d(t) - p_{s}^c(t) ] + \sum_{r \in R_z } p_r(t) + \sum_{h \in H_z} p_h(t) \\ + \sum_{l \in L} \lambda_{l,z} f_l(t) = \mathcal{L}_z(t), \; t \in T, z \in Z.
\end{multline}

\subsubsection{Unit Commitment Objective}
The unit commitment objective function \eqref{obj} is to minimize the startup and shutdown costs, fuel costs, transmission costs, and renewable curtailment costs as:
\begin{equation} \label{obj}
    \min \mathcal{C}^{gen} ;
\end{equation}
\begin{multline}
\mathcal{C}^{gen} = \sum_{t \in T} \sum_{u \in U} \Big\{SUC_u(t)+SDC_u(t) \\ + (GCI_u \cdot v_u(t) + GCS_u \cdot p_u(t)) \times 1 \: hour \Big\} \\  +  \left[\sum_{t\in T} \sum_{l \in L}  f_l(t) \cdot \emph{c}^{tx}_l + \sum_{t\in T} \sum_{r \in R} \emph{c}^{curt}_r \cdot p_r^{curt}(t)\right] \times {1 \: hour}. \nonumber
\end{multline}

\subsection{Decarbonization Planning} \label{sec3B}
The objective of decarbonization planning is to minimize the total cost associated with achieving net-zero carbon emissions from power generation by 2045. The total cost encompasses both annual energy expenses (including maintenance) and the capital costs of constructing new capacity of zero-carbon resources and lower-carbon power plants.

In the present study, it is assumed that the development of new resources will be restricted to the CAISO territory. However, the problem formulation remains broadly applicable. The portfolio of potential resources encompasses wind, solar, and energy storage at various sites, as well as geothermal, biomass, and several categories of gas-fired power plants. Decisions regarding the retirement of existing thermal units may also be considered, with certain technologies, such as coal and nuclear, already possessing predetermined decommissioning schedules. In this section, all constraints will be enforced for each year $\forall y \in Y$, and for CAISO only $z=0$ where zonal subsets of resources are concerned.

First, let us define the build status of thermal units. Let $IU_{u}(y)$ represent the binary operational status of unit $u$ in year $y$, where 1 indicates the unit is operational and may be turned on. $IU_u^{p}(y)$ denotes the planned status of unit $u$, where 1 signifies the unit is operational, and 0 indicates the unit is decommissioned or not yet constructed. $IU_u^b(y)$ and $IU_u^r(y)$ define whether the unit is built and retired, respectively, in year $y$. Consequently, the relationship between the planning layer and the unit commitment is expressed in \eqref{OUV}, which constrains the unit commitment status $v_u$ to turn on only if it is operational as defined by \eqref{IC}.
\begin{equation} \label{IC}
\begin{split}
    IU_u(y) = IU_u^p(y) + \sum_{\mathcal{Y}=1}^{y} ( IU_u^b(\mathcal{Y}) - IU_u^r(\mathcal{Y}))
\end{split}
\end{equation}
\begin{equation} \label{OUV}
\begin{split}
    IU_u(y) \geq v_{u}(y,w,t),\; \forall u \in U,\: w \in W,\: t \in T
\end{split}
\end{equation}

Installation of additional capacity of renewable generation units (indexed by $r$) and storage units (indexed by $s$) is considered to be a continuous variable and the logic capturing the installation capacity follows that of \eqref{IC} as: 
\begin{flalign} \label{ICE}
    IC_s(y) = IC_s^p(y) + \sum_{\mathcal{Y}=1}^{y} ( IC_s^b(\mathcal{Y}) - IC_s^r(\mathcal{Y})); \\
    ICE_s(y) = ICE_s^p(y) + \sum_{\mathcal{Y}=1}^{y} ( ICE_s^b(\mathcal{Y}) - ICE_s^r(\mathcal{Y})).
\end{flalign}

\noindent New capacity of these types can be installed in discrete amounts on the order of tens of watts, which is effectively continuous compared to the scale at which these resources are installed. Storage capacity has two components, one each for energy capacity (MWh), denoted as $ICE$, and power capacity (MW), denoted as $IC$. The total installed capacity of each resource is defined in a similar way to that of the thermal units, with the chief difference being the decision variables become continuous instead of binary:
\begin{equation} \label{ICR}
\begin{split}
    IC_r(y) = IC_r^p(y) + \sum_{\mathcal{Y}=1}^{y} ( IC_r^b(\mathcal{Y}) - IC_r^r(\mathcal{Y})).
\end{split}
\end{equation}
The installed capacities of these units impact the unit commitment formulation in different ways. The maximum rate of charge/discharge is equal to the rated capacity, represented as $\overline{p}_s^{c}(y)  = \overline{p}_s^d(y)  = IC_s(y)$. The maximum/minimum state of charge corresponds to the rated energy capacity, multiplied by a percentage factor associated with the operational range, denoted as $\overline{C}_s(y)=ICE_s(y) \cdot \epsilon^{max}_s$. For batteries, these values typically range between $0.1$ and $0.9$ for degradation considerations \cite{batteryDOD}, while for pumped storage, they are closer to $0$ and $1$ \cite{pumpDOD}. Regarding renewables, $IC_r(y)$ defines $IC_r$ for the specified year in \eqref{renpower}.

Let the cost of energy generation for year $y$ be denoted as $\mathcal{C}_{y}^{gen}$. This cost is composed of the same components as the function being minimized in the unit commitment given by \eqref{obj}. Within the planning problem, unit commitment is performed for a sample of several weeks per year. Each sampled week is assigned a weight $\omega_w$ that conveys its representative factor to the annual load profile, with the sum of these weights amounting to 52, corresponding to the number of weeks in a year. The yearly unit commitment cost is calculated as the weighted sum of the weekly unit commitment costs. It is also weighted by the yearly weight $\omega_y$, which encodes the number of years represented by $y$.
Consequently, the cost of generation in year $y$ and week $w$ in \eqref{obj} is expressed as $C^{gen}_{y,w}$, and the annual generation costs can be written as in \eqref{yearlygen} as: 
\begin{equation} \label{yearlygen}
\begin{split}
    \mathcal{C}_y^{gen} = \omega_y\sum_{w \in W} \omega_w \cdot{C}_{y, w}^{gen}.
\end{split}
\end{equation}
Yearly maintenance costs are considered as a function of the installed capacity and the cost of maintaining a given technology. Renewables have a single cost component expressed in \$/MW. Thermal units have a single cost component in \$/unit. Storage has two maintenance cost components, for rated energy $c^m_{s,E}$ and rated power $c^m_{s,P}$, expressed in \$/MWh and \$/MW, respectively. The cost of maintenance for the year $y$ is then: 
\begin{flalign}
 \mathcal{C}_y^{m}&= \omega_y \cdot  \biggr(\sum_{u \in U} IU_{u, y} \cdot c^{m}_i + \sum_{s \in S} ICE_{s, y} \cdot c^{m, E}_{s} \\ & + \sum_{s \in S} IC_{s, y} \cdot c^{m, P}_{s} + \sum_{k \in K} IC_{k, y} c^{m}_k + \sum_{h \in H} IC_{h, y} \cdot c^{m}_h \biggr). \nonumber
\end{flalign}
Lastly, let us consider the investment costs for constructing new resources. Annualized costs are assessed for every year after a resource is constructed. Each thermal technology is associated with an annualized capital cost per unit, denoted as $c_{y,u}^{cap}$. Similarly, storage and renewable technologies have an annualized capital cost per megawatt (\$/MW), represented as $c_{y,s}^{cap, P}$ and $c_{y,r}^{cap}$, respectively. Additionally, storage has an annualized capital cost for energy capacity, expressed as $c_{y,s}^{cap, E}$: 
\begin{flalign} \label{invcosts}
    \mathcal{C}_y^{inv} & =  \sum_{\gamma=y}^{|Y|} \omega_\gamma \cdot  \biggr(\sum_{u \in U} (IU_u^b(y)) \cdot c_{y,s}^{cap} + \sum_{s \in S} (IC_s^b(y)) \cdot c^{cap, P}_s  \nonumber \\ & + \sum_{s \in S} (ICE_s^b(y)) \cdot c^{cap, E}_s  + \sum_{r \in R} (IC_r^b(y)) \cdot c^{cap}_r\biggr).
\end{flalign}

The objective function of decarbonization planning is:
\begin{equation}
    \min \mathbb{O} = \min \sum_{y \in Y} \big\{\mathcal{C}_y^{gen} + \mathcal{C}_y^{m} + \mathcal{C}_y^{inv}\big\}.
\end{equation}
The planning process in decarbonization is subject to various constraints, which are central to formulating effective strategies. These constraints comprise emissions targets, renewable energy penetration, and system reliability. In this study, the analysis focuses solely on the constraints utilized by CAISO, excluding other balancing authorities. Thus, for notational clarity, the subscript $z$ denoting zone will be used, with the specification that $z=0$ for these planning constraints. 

Carbon emissions are generated when energy is produced by thermal plants. CAISO is subject to an emissions constraint specifying that the emissions associated with all generation within CAISO, as well as emissions associated with imports, must be less than the emissions target for year $y$, $E_y$. Given the emissions associated with unit $u$ in tons/MW as $e_u$, and the emissions associated with imports $e_l$, we can examine the relationship between emissions and energy generation. 
\begin{flalign} \label{emissions}
    E_y \geq &\sum_{w \in W} \omega_w \cdot \sum_{t \in T} \biggl(\sum_{u \in U_{z}} e_u \cdot p_u(y,w,t) \nonumber\\ & + \sum_{l \in L} e_l \cdot \max (0, \lambda_{l,z}f_l(y, w, t))  \biggr)
\end{flalign}

Only imports count towards the emission constraint, and exports should not count to reduce the emissions. Thus, the contributing emissions are lower-bounded by 0.

In addition to the net-zero emissions target by 2045, renewable portfolio standards (RPS) impose interim requirements on the proportion of electricity generation from carbon-free resources, based on a percentage of CAISO's annual load. The majority of renewable energy sources qualify for RPS, with the notable exceptions of combined heat and power (CHP) and nuclear power, which are categorized alongside renewables due to their similar generation attributes. The following constraint ensures that the renewable portfolio standards (RPS) are met for each year $y \in Y$ as:
\begin{flalign} \label{RPS}
   & RPS_y \cdot \sum_{w \in W} \sum_{t \in T} \mathcal{L}_z(y, w, t)  \leq \\ 
   &  \sum_{w \in W} \omega_w \cdot \sum_{r \in R} \sum_{t \in T} p_r(y,w,t) \cdot RPS^{eligible}_r. \nonumber
\end{flalign}
where binary variable $RPS^{eligible}_r$ indicates whether a renewable source in the set $R$ meets the RPS criteria. Accordingly, $RPS_y \cdot \sum_{w \in W} \sum_{t \in T} \mathcal{L}_0(y, w, t)$ represents the required amount of electricity generation from eligible renewable resources for the year $y$, based on a percentage ($RPS_y$) of the total annual load in CAISO.

CAISO must also satisfy reliability requirements, particularly the planning reserve margin (PRM). These requirements ensure that the portfolio, even with high renewable penetration, can meet energy demands. The PRM guarantees that the peak load, with some additional headroom, in a given year is satisfied by the installed capacity. Each resource contributes to the PRM by a fraction of its installed capacity. 

\begin{figure}
\centering
\includegraphics[width=\linewidth, trim={0 2.5cm 0 2.5cm},clip]{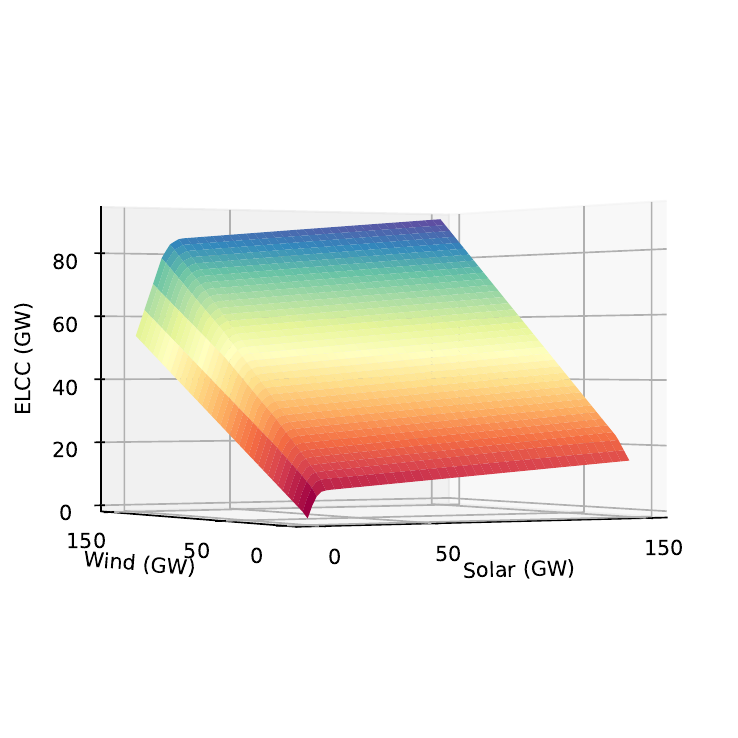}
\caption{Visualization of the ELCC surface.}
\label{fig:elcc}
\end{figure}

For thermal unit and large hydro capacity, it is modified by the net qualifying capacity fraction (NQC). Wind and solar are modified by the effective load-carrying capacity (ELCC), a fraction that decreases as renewable penetration increases. The ELCC is approximated by a 3-dimensional piece-wise linear surface, with axes representing the capacity of wind and solar, respectively. This ELCC surface is visualized in Fig. \ref{fig:elcc}. Each solar or wind resource in CAISO contributes to the total axis value by its capacity and a multiplier, denoted as $mult^{axis}$. 

The resulting 3D surface is comprised of flat segments referred to as facets, which together create the overall shape of the ELCC representation. The facets simplify the complex relationships between wind, solar, and ELCC by breaking down the surface into a series of linear segments, making it easier to analyze and understand the impact of different resource capacities on the overall system \cite{RESOLVE}.
Then, the ELCC value of each facet 
is characterized by an intercept and slope on each axis. In the optimization, the final value for this piece-wise linear ELCC is determined by setting the ELCC as the minimum of each facet value. 
With $R_{wind}$ and $R_{solar}$ denoting the subsets of CAISO wind and solar resources, the following equation computes the ELCC for each year $y$ based on the capacity of wind and solar resources in CAISO:
\begin{flalign}
ELCC_{y} \leq &\left(\sum^{r \in R_{z, wind}} IC_{y, r} \cdot mult^{axis}_{y, r}\right) \cdot slope_{y, wind, f} \nonumber \\
& + \left(\sum^{r \in R_{z, solar}} IC_{y, r} \cdot mult^{axis}_{y, r}\right) \cdot slope_{y, solar, f}  \nonumber
\\ & + intercept_{y, f}, \; f \in Facets.
\end{flalign}
Similarly, storage contributes through the 4-hour capacity. The ELCC of storage resources are characterized by a two-dimensional piecewise linear surface.
\begin{multline}
ELCC_{y, s} \leq intercept_{y, s, f} \\ +
\left(\sum^{s \in S_{z}} min(IC_{y, s}, \frac{ICE_{y, s}}{4 \: hours}) \cdot mult_{y, s}\right) \cdot slope_{y, s, f},
\\ f \in Facets.
\end{multline}

The following equation ensures that the planning reserve margin (PRM) for each year $y$ is met, taking into account the contributions from different types of resources, such as thermal units, storage, wind, and solar:

\begin{flalign} \label{PRM}
 PRM_y \leq &\sum^{u \in U_{z}} IU_{y, u} \overline{P}_u NQC_{u} \nonumber +  ELCC_{y,s} \\ &+ ELCC_{y} + 
\sum^{h \in H_{z}} IC_{y, h} NQC_{h}.
\end{flalign}

\section{Solution Methodology} \label{sec4}

The problem formulated in the previous section belongs to the class of MILP problems. MILP problems suffer from \textit{combinatorial complexity} -- because of binary decision variables, as the problem size increases, the number of possible solutions increases \textit{super-linearly} thereby leading to a drastic increase in the computational effort. In this section, to efficiently solve the problem, a recent decomposition and coordination approach \cite{bragin2022surrogate} is deployed to exploit the super-linear reduction of complexity upon the decomposition and the geometric convergence potential inherent to Polyak’s step-sizing formula for the fastest coordination possible to obtain near-optimal solutions in a computationally efficient manner.

The decomposition is operationalized by relaxing coupling zonal power balance constraints \eqref{Nodalpowerbalance}. Given the additivity of the constraints and the objective function, the relaxed problem is separable into individual unit subproblems. Subproblem solutions are first coordinated through the iterative update of Lagrangian multipliers $\mathbf{\Lambda}$. After the multipliers have converged sufficiently, the primal problem is solved while fixing the majority of the binary variables to their \textit{subproblem optimal} values.
The process for solving subproblems is described next.

\noindent \textbf{Relaxed Problem.} 
After relaxing coupling constants, the relaxed problem is broken into subproblems decomposed by groups of thermal units. The full set of thermal units is split at random into subproblem groups indexed by iteration $k$. While solving each subproblem, all other thermal units' commitment status, power levels, and build status are fixed at their value in the previous iteration. Each subproblem is optimized with respect to all the variables (both discrete and continuous) associated with the group of thermal units selected. 

After relaxing zonal power balance \eqref{Nodalpowerbalance}, which couples thermal units, the relaxed problem becomes:
\begin{flalign} 
& \min \mathbb{L} =  \min_{\{\textbf{p},\textbf{v},\textbf{I}\}} 
\begin{Bmatrix}
\mathbb{O} + \mathbf{\Lambda}  \cdot \mathbf{R} +  c \cdot \big\| \mathbf{R}\big\|_1 
\end{Bmatrix},
\\ 
\nonumber & s.t., \eqref{gen1} - \eqref{lastreservereq} \; \forall y \in Y, w \in W, \nonumber \\ & \eqref{IC} - \eqref{ICR}, \eqref{emissions} - \eqref{PRM} \; \forall y \in Y, \nonumber
\end{flalign}

\noindent where $\mathbf{R} = [r_z(y,w,t), \forall z \in Z, y \in Y, w \in W, t \in T]$ is a vector of zonal power balance constraint violations across all zones and timepoints. The violation is given by $r_z(y,w,t) = \sum_{u \in U_z} p_u(y, w, t) + \sum_{s \in S_z}[ p_{s}^d(y, w, t) - p_{s}^c(y, w, t) ] + \sum_{r \in R_z } p_r(y, w, t) + \sum_{h \in H_z} p_h(y, w, t) +  \sum_{l \in L} \lambda_{l,z}f_l(y, w, t) - \mathcal{L}_z(y, w, t)$.
$\mathbf{\Lambda}$ is a vector of Lagrangian multipliers, and $c$ is a penalty coefficient acting on the absolute value of constraint violations.
For notational brevity,
let $\textbf{p}$ represent all power-related variables (including line flows), $\textbf{v}$ represent all binary commitment variables, and $\textbf{I}$ represent all investment variables.

The solution process is presented in Algorithm \ref{alg:alg1}. Subproblems are formulated by selecting $\Omega_k$ - a group of units to be optimized with respect to at iteration $k$ and by fixing decision variables collectively denoted as $\{\textbf{p},\textbf{v},\textbf{I}\}$ that do not belong to $\Omega_k$ are fixed at previously obtained values $\{\textbf{p}^{k-1},\textbf{v}^{k-1},\textbf{I}^{k-1}\}$. In particular, all units are split at random into groups, and these groups are iterated through during the subproblems. In each iteration, all non-thermal-unit variables are also solved.

After one subproblem is solved, the multipliers are updated along ``surrogate'' subgradient directions, which are violation levels of relaxed constraints, with an appropriate stepsize as follows: 
 \begin{flalign}
 \mathbf{\Lambda}^{k}  
=
\mathbf{\Lambda}^{k-1}  
+ s^k \cdot \Tilde{\mathbf{R}}^k. \label{multiplierupdate}
 \end{flalign}
 \noindent 

Penalty term $c$ acts on the absolute value of constraint violations. Care must be taken to update the penalty term. If $c$ is initialized too large or grows too quickly, it can hamper the convergence of multipliers as well as severely impact the iteration time. Further discussion of the role of the penalty term and strategies for updating it can be found in \cite{wu2021novel}.
 
 High penetration of renewable and storage resources, however, causes issues with the convergence of multipliers, as these resources can be dispatched at identical costs. The dispatch of these resources may jump between maximum and minimum, and multipliers may oscillate and overshoot optimal multipliers. 
 Consider a toy problem consisting of a single hour in which renewable generation exceeds load. Neglecting penalty term $c$, at any positive value of multiplier $\Lambda$, the optimization $\min \mathbb{L}$ will minimize $R$ by curtailing all generation. Conversely, if $\Lambda$ is any negative number, it will curtail nothing and overgenerate. Although this is a simplified example, similar behavior is exhibited for the dispatch and construction of resources. In a sense, Lagrangian multipliers are ``price signals'' and the renewable resources with similar costs respond in a similar way potentially leading to the jumping of solutions and, consequently, to zigzagging of multipliers. To alleviate the solution jumping issue, specifically, to suppress the jumping of solutions, each continuous power variable is restricted within $\Delta = 500MW$ of its value from the previous subproblem. The procedure also alleviates zigzagging since with suppressed jumping of solutions, the corresponding multiplier-updating directions tend not to change drastically (i.e., become smoother) eventually leading to a smoother update of multipliers. From the multiplier convergence perspective, to alleviate the overshooting issue, proper stepsize selection plays an important role as explained ahead. 

\noindent \textbf{Stepsize Update.} 
The step size, denoted as $s^k$, plays a pivotal role in the proposed algorithm's convergence. Following the methodology outlined in \cite{bragin2022surrogate}, the step size is computed as follows:
\begin{flalign}
& s^{k}= \zeta \cdot \gamma \cdot \frac{\Bar{q}_k-L^k}{\| \Tilde{\mathbf{R}}\|^2}. \end{flalign}

In this equation, $\zeta$ and $\gamma$ are hyperparameters that are chosen to balance the trade-off between convergence speed and algorithm stability. While in the original work of Polyak \cite{polyak1969minimization} $\gamma < 2$, which would be appropriate for standard Lagrangian relaxation that utilizes subgradient directions for multiplier update, in further ``surrogate'' extensions of LR \cite{zhao1999surrogate, bragin2022surrogate}, $\gamma < 1$. Moreover, since term $\Bar{q}_k$ represents the current overestimation of the dual value, $\zeta$ is chosen in a way to reduce stepsizes (e.g., $\zeta = 1/2$). Table \ref{tab:lrmethods} compares the hyperparameters and dual values used for the stepsize updates in the references. Since the method solves one subproblem at a time, $\gamma$ is chosen to be the reciprocal of the number of subproblems. Overall, as compared to standard LR, our method allows for a more frequent update of multipliers along smoother directions with smaller steps leading to the alleviation of the ``overshooting'' issue mentioned above. 
\begin{table*}[tb]
\caption{Comparison of characteristics of LR methods using Polyak's stepsize as well as Polyak's seminal work}
\centering
\begin{tabular}{ |c|c|c|c|c| } 
\hline
Method & Requirement to set stepsizes & Solution/Multiplier updating directions & $\zeta$ & $\gamma$ \\
 \hline
 \\[-1em]
Polyak's seminal work (1969) \cite{polyak1969minimization} & Optimal dual value ($q^*$) & Subgradient ($R$) & - & $<2$ \\
\hline
 \\[-1em]
Surrogate subgradient method \cite{zhao1999surrogate} & Optimal dual value ($q^*$) & Surrogate subgradient ($\Tilde{R}$) & - & $<1$ \\
\hline
 \\[-1em]
Surrogate ``level-based'' Lagrangian relaxation \cite{bragin2022surrogate} & W/o optimal dual value ($q^*$) & Surrogate subgradient ($\Tilde{R}$) & $<1$ & $<1$ \\
\hline
\end{tabular} \label{tab:lrmethods}
\end{table*}

The step size is initially set using an overestimation of the optimal dual value. 
The value of $\Bar{q}_k$ is not static, rather, as detailed in \cite{bragin2022surrogate}, it undergoes periodic adjustments based on a level-based resetting mechanism, which detects the lack of multiplier convergence. This resetting process is designed to lower $\Bar{q}_k$ to approach the actual (dual) optimal value in light of new information obtained during the iterative procedure, thereby refining our overestimation of the dual value and guiding the algorithm toward the optimal multipliers. In essence, this approach to updating the step size is grounded in dynamism and adaptivity. By making use of gathered information and tuning the step size accordingly, we can expedite convergence and enhance the efficiency and robustness of the overall algorithm. For more information, please refer to \cite{bragin2022surrogate}. 

\begin{algorithm}
\SetAlgoLined
 initialize $\mathbf{\Lambda}, c, \bar{q}_0, \zeta, \gamma$

\While{$ \| \mathbf{\Tilde{R}} \| > threshold$}{
select subproblem units $\Omega_k$ \;
solve subproblem $min \, \mathbb{L}^k$, obtaining $\mathbf{\Tilde{R}}$ \;
check level $\bar{q}$ for convergence of $\mathbf{\Lambda}$ \;
update $s^k$ \;
update $\mathbf{\Lambda}^k$ \;
}
\caption{Surrogate ``Level-Based'' Lagrangian Relaxation}
\label{alg:alg1}
\end{algorithm}

\noindent \textbf{Feasible Solution.} 
Although the method is guaranteed to converge towards the optimal (dual) solution \cite{liu2023accelerating}, obtaining zero constraint violations through multipliers alone is often difficult or impossible. As a result, a heuristic is necessary to find a feasible solution to the primal problem. Once the multipliers have converged such that the constraint violations are sufficiently low, the primal problem is solved by fixing the commitment status of most units to the values obtained in the relaxed problem. The above heuristic presents a fundamental trade-off: constraining fewer variables in the primal problem requires greater computational effort but may lead to lower overall costs. Nevertheless, as empirical evidence suggests in the section ahead, by solving the entire primal problem but with respect to only a small number of units, 
the primal problem is much easier to solve and can generally be solved to near-optimality.

\section{Numerical Study} \label{sec5}
\subsection{Experimental Setup}
The decarbonization model is based on the Western Interconnection. The model incorporates CAISO and 5 other zones: LADWP, BANC, IID, and aggregations of non-California balancing authorities in the Northwest and Southwest. The data used within the model is taken from the RESOLVE implementation published by CPUC \cite{RESOLVE}. We model 8 weeks per year, and biennially from 2023 to 2045. The model has 91.1 million continuous variables and 11.8 million binary variables, roughly 40x more total variables than in RESOLVE. We use Gurobi on a workstation with an AMD Ryzen Threadripper 3970X  CPU to solve the subproblems and primal problem. The total solution process takes less than 48 hours. Each iteration accounts for approximately 20 minutes, although the exact time varies considerably, and a feasible solution can be attained after 120 iterations.

To reduce computational complexity, rather than modeling unit commitment for full years, representative weeks are sampled and used instead. The sampling method follows the method used in RESOLVE studies \cite{E3SAMPLING}. We diverge from RESOLVE by modeling representative weeks rather than representative days, and the advantages will be demonstrated in the results. 
Histogram bins $b \in B$ are created from features of the data, most importantly the distributions of hourly loads. Then, an optimization problem is solved to select weeks and corresponding weights which minimizes the Manhattan distance of bin frequency in the full year to the representative weeks. 
\begin{equation}
\begin{split}  
\min & \sum_{b \in B} \big{(} YearlyFreq_b \\ - & \sum_{w \in weeks} \omega_w \cdot WeeklyFreq_{w,b}\big{)}
\end{split}
\end{equation}

The optimization horizon is through 2045, and financing through 2065. Due to the sampling of weeks and years, it is necessary to weight the weekly and yearly components to ensure fair balance between these different cost components in the objective function. The capital costs in \eqref{invcosts} are amortized. Then, these costs are assessed for every subsequent year to the decision in the optimization horizon. Note that this also accounts for residual value of new capacity. All costs are in real dollars, but the yearly weight should capture the time value of money, with an assumed discount rate of 5\%, and the number of real years represented by the sampled year:
\begin{equation}
\begin{split}
    \omega_y = \left(\frac{1}{1.05}\right)^{(Y_y-Y_0)} \cdot (Y_{(y+1)} - Y_{y}).
\end{split}
\end{equation}

Within the objective function \eqref{obj}, each yearly component term is weighted by its respective yearly weight in the case of investment and maintenance costs (computed yearly) and weighted by both yearly weight and weekly weight in the case of generation costs. For year 2045, the yearly weight captures operation through the year 2065. 

\subsection{Results}

We compare the fleet of RESOLVE to the fleets of our model when optimized over both the same representative days as RESOLVE, and the proposed representative weeks. Operating cost comparisons are obtained by fixing investment decisions and solving for $C_y^{gen}$. Cost comparisons are presented using both the proposed representative weeks and RESOLVE's representative days. The proposed method with representative weeks will be generally referred to as MILP + LR, with the suffix of ``day'' or ``week'' added where the distinction is necessary.

\begin{figure}[htb]
\centering
\includegraphics[width=1\linewidth]{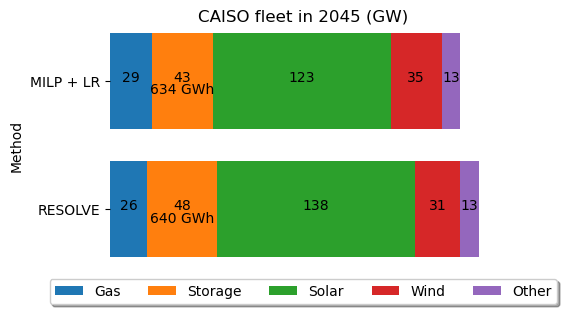}
\caption{Comparison of CAISO fleet in 2045.}
\label{fig:fleet}
\end{figure}

\begin{figure*}
\centering
\includegraphics[width=0.80\textwidth]{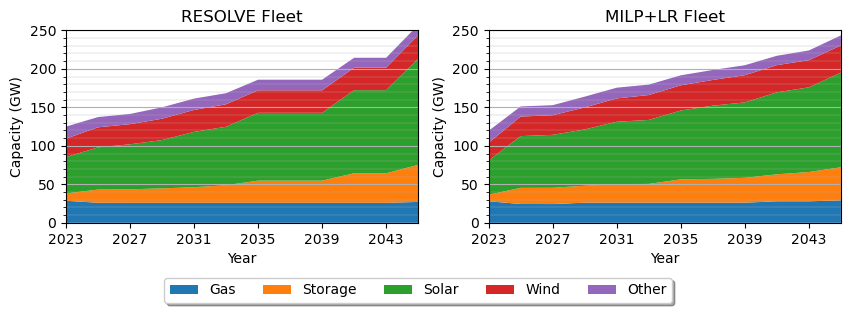}
\caption{Comparison of CAISO fleet over time.}
\label{fig:yearfleet}
\end{figure*}

With the goal of decarbonization by 2045 in mind, a comparison of CAISO fleets in 2045 is shown in Figure \ref{fig:fleet}. Compared to RESOLVE, our model builds fewer resources in all categories except wind. However, this is not the case throughout the optimization horizon. Figure \ref{fig:yearfleet} shows the fleet composition from 2023 to 2045. Our model begins investing in additional capacity, especially solar, much earlier than RESOLVE, and holds more total capacity than RESOLVE until 2045. 

The likely explanation for the lower build of renewable resources in RESOLVE during this transitional years is that the overestimated flexibility of thermal resources underestimates the running costs of thermal units, as well as their emissions. In fact, RESOLVE's investment decisions, when applied to the MILP+LR model, cannot satisfy emissions constraints between year 2027 and 2045, and overemits anywhere from a few thousand to several million tons of GHG per year. Due to this underestimation, RESOLVE chooses to defer investment in renewables to later years. As evidence of the cost underestimation of running gas generators, Table \ref{tab:baselinecosts} shows the shutdown, startup, and fuel costs for CAISO gas generators in 2022, before substantial investment occurs. The costs shown for RESOLVE are with respect to their linearized, clustered unit commitment formulation. RESOLVE underestimates fuel costs by roughly 20\% and, vastly underestimates startup and shutdown costs. Thus, this may explain why RESOLVE's investment plan leans more heavily on gas units. Similarly, RESOLVE drastically underestimates the emissions in 2022, although it is still within emissions limits. California Air Resources Board estimates emissions of roughly 40MMT from in-state electricity production in 2020, which is much more aligned with our model than RESOLVE. The takeaway is that if RESOLVE results are used to inform policy decisions, it may be difficult to meet intermediate emissions targets due to under-investment in renewable energy and storage. 

By a similar token, the differences in wind vs solar investment may be explained by the more accurate modeling of gas generators. When gas generators are modeled more faithfully, wind may have a more complementary load shape. Even today, ramping poses difficulties during early evening hours in which load is increasing and solar generation is rapidly decreasing. With solar capacity more than doubled, this effect will become even more pronounced. 

\begin{table}[H]
\caption{Comparison of CAISO 2022 Baseline Generation Costs}
\centering
\begin{tabular}{ |c|c|c|c|c| } 
\hline
 & Shutdown & Startup & Fuel & Emissions \\
 & (Millions \$) & (Millions \$) & (Millions \$) & (MMT) \\
 \hline
MILP + LR & 29.71  & 70.46 & 2594.53 & 35.7 \\
\hline
RESOLVE & 5.28 & 5.28 & 1935.45 & 21.4 \\
\hline
\end{tabular} \label{tab:baselinecosts}

\end{table}

\begin{figure}[htb]
\centering
\includegraphics[width=.92\linewidth]{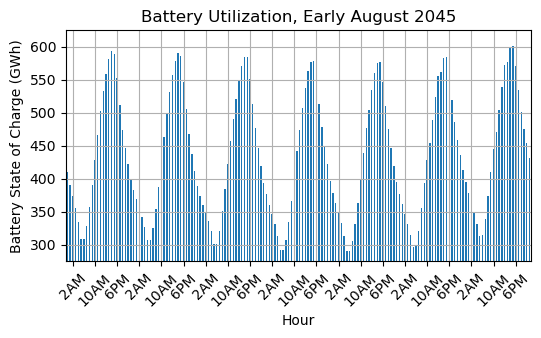}
\caption{Battery state of charge for an exemplary week in 2045.}
\label{fig:battery}
\end{figure}

The storage state of charge for an exemplary week in August 2045 is shown in Fig \ref{fig:battery}. As expected, the state of charge is at its maximum in the late afternoon and its minimum in the early morning. A feature of note is that the state of charge maximum and minimum vary by roughly 30GWh. This indicates that energy is being shared between days, behavior which is enabled by modeling longer representative periods. In contrast, if energy sharing is not allowed, as in the case of modeled days in RESOLVE, the state of charge must be equal at the beginning and end of the day. Also of note is that our model builds 5 less GW of storage, but only 6GWh less energy capacity. In 2045, our model builds battery storage with approximately 8 hour duration vs 7 hour in RESOLVE, and pumped storage with 100 hour duration vs 90 hour in RESOLVE.

\begin{figure}[htb]
\centering
\includegraphics[width=\linewidth]{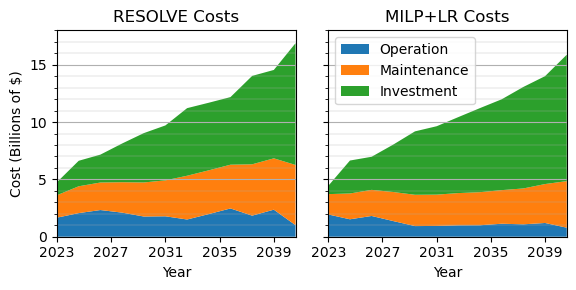}
\caption{Yearly cost breakdown for CAISO.}
\label{fig:yearcosts}
\end{figure}

Yearly costs are shown in Fig \ref{fig:yearcosts}. Our model maintains lower operation costs in almost every year. Investment spending is higher in early years, but increases at an overall lower rate, resulting in lower investment costs in the second half of the study period. Costs shown in Fig \ref{fig:yearcosts} are yearly, including financing of investment from earlier years, and not adjusted for discount rate. As shown in Table \ref{tab:Total Costs}, with generation, investment, and maintenance costs included but neglecting emission violations, our model presents a savings of $1.2 \%$, or 4 billion dollars through 2065. This includes all costs for CAISO as well as operating costs for the other WECC zones. As a reminder, investment in non-CAISO zones is exogenous, and CAISO investment has a limited impact on outside operating costs. When considering CAISO costs alone, our model saves 1.9\%. 
\begin{table}[H]
\caption{Comparison of Total System Costs \\ 2022 \$, Billions}
\centering
\begin{tabular}{ |c|c|c|c|c|c| } 
\hline
 & System Cost & \multicolumn{4}{c|}{CAISO Costs}\\
 \hline
 & Total & Total & Op. & Maint. & Invest. \\
 \hline
RESOLVE & 341.35 & 205.86 & 32.42 & 65.38 & 108.06 \\ 
\hline
MILP + LR, Day & 336.47 & 202.53 & 26.23 & 52.96 & 123.35 \\
\hline
MILP + LR, Week & 337.36 & 201.93 & 22.73 & 56.01 & 123.19 \\
\hline
\end{tabular} \label{tab:Total Costs}
\end{table}

\begin{table}[H]
\caption{Comparison of total CAISO costs under range of per ton carbon costs. \\ 2022 \$, Billions}
\centering
\begin{tabular}{ |c|c|c|c| } 
\hline
 & \multicolumn{3}{c|}{Total CAISO Costs}\\
 \hline
 Tax & \$0 & \$30 & \$100 \\
 \hline
RESOLVE & 205.86 & 207.514 & 211.38 \\ 
\hline
MILP + LR, Day & 202.53 & 203.07 & 204.31  \\
\hline
MILP + LR, Week & 201.93 & 201.93 & 201.93  \\
\hline
\end{tabular} \label{tab:emis_costs}
\end{table}

Table \ref{tab:emis_costs} shows the total CAISO costs under different values of a carbon tax per ton of GHG emissions over the limit. A carbon tax of \$30 roughly corresponds to the 2022 California cap-and-trade clearing price \cite{captrade}. A carbon tax of \$100 roughly corresponds to the Department of Energy's Carbon Negative Shot goal for direct air capture cost per ton \cite{earthshot}. These two carbon taxes essentially bookend the cost of exceeding emission targets. Including some component for the cost of emissions is critical, as exceeding emissions targets comes at the direct benefit of avoided investment in renewables and storage. 
Due to sampling days vs weeks and increasing the total number of modeled days roughly 50\% from 37 to 56, the average load in the representative weeks scenario is slightly higher than the representative days. Emissions constraints are binding constraints in most years in all three models. Thus, it is unsurprising that our own MILP + LR Day investment decisions have emissions violations in the representative week evaluation. Crucially, the MILP + LR Day scenario has less than 1/3rd of the total violation of RESOLVE.

The comparison of RESOLVE to MILP+LR over representative days provides the most isolated comparison of the value of more rigorous generator modeling. The results of this experiment considering a range of carbon taxes are presented in Table \ref{tab:daycosts}. In this scenario, again our model produces no violation of emissions, while RESOLVE investments result in several million tons of GHG overemitted each year. Even neglecting any cost of overemission, our more detailed model has a nearly 5\% lower total cost, primarily due to increased investment and correspondingly lower operating costs. With a \$100 carbon cost, the gap grows to over 7\%. Again, the message is the same: the simplifications to generator modeling underestimate both emissions and fuel costs. This both underestimates the requirement for and value of renewable and storage resources.

\begin{table}[H]
\caption{Comparison of total CAISO costs with representative days under range of per ton carbon costs. \\ 2022 \$, Billions}
\centering
\begin{tabular}{ |c|c|c|c| } 
\hline
 & \multicolumn{3}{c|}{Total CAISO Costs}\\
 \hline
 Tax & \$0 & \$30 & \$100 \\
 \hline
RESOLVE & 204.34 & 205.96 & 209.73 \\ 
\hline
MILP + LR, Day & 194.84 & 194.84 & 194.84  \\
\hline
\end{tabular} \label{tab:daycosts}

\end{table}

\section{Conclusion}
In this paper, a MILP decarbonization model for California is developed. To overcome the issue of combinatorial complexity with integer variables, SLBLR is implemented, and allows us to optimize over nearly 100 million variables including 12 million binary variables in under 48 hours. We show that the existing, linearized model underestimates operational costs of gas generators, leading to a substantially different investment plan. By doing so, we develop an investment plan that saves California 4 billion dollars over the investment horizon. Further, our model suggests more substantial and early investment in renewable generation and storage is required to meet intermediate emissions targets. This result may inform policymakers that a more aggressive approach is needed than previous work sponsored by state commissions. 
Future works will use this model to investigate the impacts of external factors on the optimal decarbonization pathway, such as climate change and vehicle-to-grid charging. 

\bibliographystyle{ieeetr}
\bibliography{references}

\end{document}